\documentclass[10pt,english,british,tightenlines,eqsecnum,
floats,aps,amsmath,amssymb,nofootinbib,superscriptaddress,prd,showpacs,showkeys]
              {revtex4}
\usepackage[latin9]{inputenc}
\usepackage{color}
\usepackage{babel}
\usepackage{amsmath}
\usepackage{amssymb}
\usepackage{graphicx}
\usepackage{esint}
\usepackage[unicode=true,pdfusetitle,
 bookmarks=true,bookmarksnumbered=false,bookmarksopen=false,
 breaklinks=false,pdfborder={0 0 1},backref=false,colorlinks=false]
 {hyperref}

\makeatletter
\@ifundefined{textcolor}{}
{%
 \definecolor{BLACK}{gray}{0}
 \definecolor{WHITE}{gray}{1}
 \definecolor{RED}{rgb}{1,0,0}
 \definecolor{GREEN}{rgb}{0,1,0}
 \definecolor{BLUE}{rgb}{0,0,1}
 \definecolor{CYAN}{cmyk}{1,0,0,0}
 \definecolor{MAGENTA}{cmyk}{0,1,0,0}
 \definecolor{YELLOW}{cmyk}{0,0,1,0}
}

\usepackage{babel}

\usepackage{euscript}\usepackage{epsfig}

\usepackage{amsfonts}

\usepackage{fancyhdr}

\makeatother

\begin{document}
%-------------------------------------------------------------------------------
\title{\bf Breaking the Democratic Limit in the Generalized Friedberg-Lee Model: Implications for Neutrino Masses and Mixing }
% Neutrino Mass Matrix and Flavor Mixing within a Broken Democratic Limit of the Generalized Friedberg�Lee Model

\author{N. Razzaghi}
\email{n.razzaghi@iau.ac.ir}

\affiliation{Department of Physics, QA.C., Islamic Azad
University, Qazvin, Iran}

\begin{abstract}

We investigate a generalized Friedberg--Lee (FL) framework for
neutrino masses, focusing on the singular parameter space at Point
D ($\alpha = \beta = -1/3$). At this limit, the neutrino mass
matrix exhibits a democratic texture governed by the $S_3$
permutation symmetry. Although theoretically profound, the exact
democratic limit is phenomenologically excluded as it predicts a
degenerate mass spectrum and a vanishing reactor angle
($\theta_{13}=0$). To reconcile this high-symmetry limit with
experimental observations, we introduce a minimal and systematic
perturbation that preserves the Twisted Friedberg--Lee (TFL)
symmetry. This mechanism effectively lifts the mass degeneracy and
breaks the magic and $\mu$--$\tau$ symmetries in a controlled
manner. Our derivation avoids \textit{ad hoc} parameters,
establishing a robust framework that yields a realistic inverted
mass hierarchy ($m_3=0$), a non-zero $\theta_{13}$, and intrinsic
CP violation. We demonstrate that the model predicts maximal
atmospheric mixing and a significant Dirac CP phase. The obtained
numerical ranges for the neutrino masses and the Jarlskog
invariant show excellent agreement with global fit data, providing
a theoretically motivated foundation for neutrino flavor physics
within the TFL scheme.

\end{abstract}


\date{July 19, 2026}

%\pacs{02.40.Gh, 04.70.Bw, 04.20.Dw}

\maketitle

%===============================================================================
%===============================================================================
\section{Introduction}\label{sec1}
%===============================================================================
%===============================================================================

The non-zero nature of neutrino masses has been robustly
established by a wealth of oscillation data \cite{exp, exp1}.
Based on the most recent global fits, the current experimental
landscape for neutrino oscillation parameters at the $3\sigma$
confidence level is summarized as follows \cite{newexp}:

\begin{eqnarray}\label{exp}
\delta m^{2}[10^{-5}\text{eV}^{2}]&=&(6.92-8.05),\nonumber\\
|\Delta m^{2}|[10^{-3}\text{eV}^{2}]&=&(2.463-2.606)-(2.438-2.584),\nonumber\\
\sin^{2}\theta_{12}&=&(0.275-0.345),\nonumber\\
\sin^{2}\theta_{23}&=&(0.430-0.596)-(0.437-0.597),\nonumber\\
\sin^{2}\theta_{13}&=&(0.02023-0.02376)-(0.02053-0.02397),\nonumber\\
\delta&=&(96^\circ-422^\circ)-(201^\circ-348^\circ),
\end{eqnarray}

The provided ranges are presented in two columns: the left column
corresponds to the normal hierarchy (NH), while the right column
corresponds to the inverted hierarchy (IH), where $\delta
m^2\equiv m_2^2-m_1^2$ and $\Delta m^2\equiv m_3^2-m_1^2$.

Experimental measurements have confirmed that $\theta_{13}$ is
non-zero with more than $5\sigma$ significance \cite{exp1},
although it remains small relative to the other neutrino mixing
angles.

The lepton mixing matrix is conventionally parameterized as
\cite{mixing}:
\begin{equation}\label{emixing}
U_{PMNS}=\left(\begin{array}{ccc}c_{12}c_{13} & s_{12}c_{13} & s_{13}e^{-i\delta}\\
-s_{12}c_{23}-c_{12}s_{23}s_{13}e^{i\delta} &
c_{12}c_{23}-s_{12}s_{23}s_{13}e^{i\delta} &
s_{23}c_{13}\\s_{12}s_{23}-c_{12}c_{23}s_{13}e^{i\delta}
& -c_{12}s_{23}-s_{12}c_{23}s_{13}e^{i\delta} & c_{23}c_{13}\end{array}\right)\left(\begin{array}{ccc}e^{i\rho}  & 0 & 0 \\
0 & 1& 0\\0 & 0 & e^{i\sigma}\end{array}\right),
\end{equation}
where $c_{ij}\equiv\cos\theta_{ij}$ and
$s_{ij}\equiv\sin\theta_{ij}$ for $i,j=1,2,3 \hspace{2mm} (i<j)$.
Here, $\delta$ represents the Dirac CP violating phase, which is
analogous to the CKM phase, while $\rho$ and $\sigma$ denote the
Majorana phases that are applicable if neutrinos are Majorana
particles.

The Dirac mass Lagrangian for neutrinos and charged leptons can be
expressed as:
 \begin{equation}\label{e2}\vspace{.2cm}
  {\cal L_\mathbf{m}}= {-\bar{\ell}_{{Li}}}{\cal
M_\mathbf{e}}^{ij}\ell_{Rj}-{\bar{\nu}_{{Li}}}{\cal
M_\mathbf{D}}^{ij}\nu_{Rj}+h.c.,
\end{equation}

Friedberg and Lee (FL) proposed a successful phenomenological
model for neutrino masses \cite{FL}, which is characterized by an
appropriate flavor symmetry for Dirac neutrinos. Within this
framework, the charged lepton mass matrix is diagonal.
Consequently, the neutrino mixing is uniquely described by a
$3\times3$ unitary matrix $U$ that maps the neutrino mass
eigenstates onto the flavor eigenstates
$(\nu_{e},\nu_{\mu},\nu_{\tau}).$ In the limit of the pure FL
model, one of the neutrino mass eigenvalues vanishes identically.
This feature partially accounts for the smallness of the neutrino
mass scale. Moreover, in the presence of $\mu-\tau$ symmetry, the
mixing matrix $U$ simplifies to the tribimaximal (TBM) mixing
pattern \cite{TBM}.

The Dirac neutrino mass operator within the FL framework is given
by:
\begin{eqnarray}\label{e3}\vspace{.5cm}
  {\cal M_\mathbf{FL}}&=&a\left(\bar{\nu}_{\tau}-\bar{\nu}_{\mu}\right)\left(\nu_{\tau}-\nu_{\mu}\right)
+ b\left(\bar{\nu}_{\mu}-\bar{\nu}_{e}\right)\left(\nu_{\mu}-\nu_{e}\right)\nonumber\\
&+&
c\left(\bar{\nu}_{e}-\bar{\nu}_{\tau}\right)\left(\nu_{e}-\nu_{\tau}\right)
+
m_{0}\left(\bar{\nu}_{e}\nu_{e}+\bar{\nu}_{\mu}\nu_{\mu}+\bar{\nu}_{\tau}\nu_{\tau}\right).
\end{eqnarray}

All parameters in this model ($a, b, c$, and $m_{0}$) are assumed
to be real valued. In the limit $m_0 \to 0$, the Lagrangian
exhibits a shift symmetry: $\nu_{e}\rightarrow\nu_{e}+z$,
$\nu_{\mu}\rightarrow\nu_{\mu}+z$, and
$\nu_{\tau}\rightarrow\nu_{\tau}+z$, where $z$ is an element of
the Grassmann algebra. In the specific case where $z$ is a
constant, the Friedberg Lee (FL) symmetry \cite{FL} is recovered,
ensuring the invariance of the kinetic term. However, the presence
of the $m_{0}$ term introduces an explicit breaking of this
symmetry within the electroweak Lagrangian.

Notably, the FL symmetry gives rise to a magic matrix structure,
which remains preserved even in the presence of the $m_{0}$ term.
The various manifestations of this magic symmetry are explored in
detail in the subsequent sections. Furthermore, the FL symmetry
has been identified as the residual symmetry of the neutrino mass
matrix emerging from the breaking of $SO(3)\times U(1)$ flavor
symmetry \cite{FL2}.

The structure of the resulting mass matrix is given by:
 \begin{equation}\label{e4}
M_{FL} =\left(\begin{array}{ccc}b+c+m_{0} & -b & -c\\
-b & a+b+m_{0} & -a\\-c & -a & a+c+m_{0}\end{array}\right),
\end{equation}
where the parameters are related to the Yukawa coupling elements
$Y_{\alpha\beta}$ via $a \propto (Y_{\mu\tau}+Y_{\tau\mu})$, $b
\propto (Y_{e\mu}+Y_{\mu e})$, and $c \propto (Y_{\tau
e}+Y_{e\tau})$ \footnote{The proportionality constant is the Higgs
vacuum expectation value (VEV).}. The matrix $M_{FL}$ possesses an
exact $\mu-\tau$ symmetry provided that $b=c$. Under this
condition, and assuming the hermiticity of $M_{FL}$, the mass
matrix is diagonalized by the TBM matrix as
$U^{T}_{TBM}M_{FL}U_{TBM}=\text{Diag}\{m_{1},m_{2},m_{3}\}$,
yielding the following mass eigenvalues:
\begin{equation}\label{e6}\vspace{.2cm}
m_{1}=3b+m_{0}, \quad m_{2}=m_{0}, \quad m_{3}=2a+b+m_{0}.
\end{equation}
This result reproduces the well known tribimaximal (TBM) mixing
pattern \cite{TBM}, which is characterized by the matrix:
\begin{equation}\label{etbm}
U_{TBM} =\left(\begin{array}{ccc}-\sqrt{\frac{2}{3}} & \frac{1}{\sqrt{3}} & 0\\
\frac{1}{\sqrt{6}} & \frac{1}{\sqrt{3}} &
-\frac{1}{\sqrt{2}}\\\frac{1}{\sqrt{6}} & \frac{1}{\sqrt{3}} &
\frac{1}{\sqrt{2}}\end{array}\right).
\end{equation}

It is well known that the exact TBM limit predicts
$(U_{e3})_{TBM}=0$, which stands in tension with current
experimental evidence of a non-zero $\theta_{13}$. A non-vanishing
$\theta_{13}$ is not only crucial for enabling CP violation in the
lepton sector (requiring both $\theta_{13}\neq0$ and
$\delta\neq0$) but also provides a necessary mechanism for
leptogenesis. Moreover, the requirement of $\theta_{13}\neq0$
aligns the lepton mixing pattern more closely with the qualitative
features of the quark sector, where multi-generational mixing is a
confirmed phenomenon, despite the significant disparity in the
magnitudes of the respective mixing angles.

From Eq. (\ref{etbm}), the neutrino mass matrix in the flavor
basis can be decomposed as:
\begin{equation}\label{em3}
{\cal{M}} =\frac{m_1}{6}\left(\begin{array}{ccc}4 & -2 & -2\\
-2 & 1 & 1\\-2 & 1 & 1\end{array}\right)+\frac{m_2}{3}\left(\begin{array}{ccc}1 & 1 & 1\\
1 & 1 & 1\\1 & 1 & 1\end{array}\right)+\frac{m_3}{2}\left(\begin{array}{ccc}0 & 0 & 0\\
0 & 1 & -1\\0 & -1 & 1\end{array}\right),
\end{equation}
where $m_i$ ($i=1,2,3$) are the neutrino mass eigenvalues.
Notably, the second term in Eq. (\ref{em3}) corresponds to the
democratic matrix structure \cite{demo1}. This structure arises
from a phenomenological model of Dirac neutrino masses predicated
on an $S_3^L\times S_3^R$ flavor symmetry. In this framework, a
democratic basis is adopted such that all entries of the mass
matrix are equal. Within this basis, the $S_3$ group operations
correspond to permutations of the three generation indices. The
invariance of the Lagrangian under these permutations necessitates
a universal coupling strength for the three generations, implying
that they transform as a triplet representation of $S_3$, while
the Higgs field transforms as an $S_3$ singlet.

The smallness of $\theta_{13}$ relative to other mixing angles
suggests that the TBM structure requires a minor perturbation to
accommodate current experimental data and yield a realistic
neutrino mixing matrix. In this context, we examine the Friedberg
Lee (FL) neutrino mass model in a specific limit that reproduces
the democratic mass matrix in the flavor basis. While the
democratic structure is theoretically appealing for generating a
non-zero $\theta_{13}$ from an initial TBM baseline, the resulting
mass spectrum (when diagonalized by the standard $U_{TBM}$)
appears to be in tension with experimental observations.
Consequently, the primary objective of this study is to derive an
experimentally viable mass spectrum and a realistic mixing matrix
by introducing small perturbations into the fundamental TBM
framework. Although various models have been proposed to achieve
the TBM form \cite{8}, and numerous studies have explored
mechanisms to generate $\theta_{13}\neq0$ from a TBM baseline
\cite{ttbm}, achieving a consistent description of CP violation,
which requires both $\delta\neq0$ and $\theta_{13}\neq0$ within
the standard parametrization of Eq. (\ref{emixing}), remains a
central challenge.

The Jarlskog rephasing invariant parameter $J$ \cite{J}, defined
as
\begin{equation}\label{eJ}\vspace{.2cm}
J={\cal{I}}m(U_{e1}U^{\ast}_{e2}U^{\ast}_{\mu 1}U_{\mu 2}),
\end{equation}
serves as a crucial indicator of CP violation in lepton number
conserving processes, such as neutrino oscillations. Notably,
neutrino oscillation experiments alone cannot distinguish between
Dirac and Majorana neutrinos \footnote{The detection of
neutrinoless double beta decay ($0\nu\beta\beta$) would provide
definitive evidence for lepton number violation and confirm the
Majorana nature of neutrinos.}. Consequently, extensive
theoretical and phenomenological research has focused on neutrino
mass models that incorporate $\mu-\tau$ symmetry breaking as a
necessary prerequisite for CP violation \cite{theoretical}.

In this paper, we generalize the FL model by introducing complex
parameters. Within this generalized framework, we impose physical
constraints to ensure that the mass eigenvalues remain real. We
then identify a confined region in the parameter space where CP
violation emerges. Of particular interest is a specific boundary
point, designated as Point D, at which the neutrino mass matrix
reduces to the democratic structure. Since the democratic matrix
is phenomenologically inconsistent with experimental data
(specifically yielding $m_1=m_3=0$), we extend the model to allow
for $m_1\neq 0$ by introducing a symmetry breaking term based on
the underlying FL symmetry.

To satisfy experimental constraints, we analyze the neutrino mass
matrix within the $U_{TBM}$ structure. This leads to an
unperturbed mass matrix in the flavor basis that preserves both
$\mu-\tau$ symmetry and the magic property. However, as this
unperturbed configuration fails to account for the solar mass
splitting, we introduce a small perturbation to generate the
required mass splitting and facilitate a non-vanishing $U_{e3}$.
By employing perturbation theory in the mass basis with complex
elements, we induce a mild breaking of the $\mu-\tau$ symmetry.
This mechanism simultaneously generates a non-zero $\theta_{13}$
and a non-trivial CP violating phase $\delta$, ultimately yielding
a realistic neutrino mixing matrix in agreement with current
observations.

The remainder of this paper is organized as follows. In Section 2,
we introduce the generalized FL model and establish the conditions
for the critical point within the expanded parameter space. We
then perform a complex perturbation analysis to derive the
perturbation mass matrix, which is responsible for generating the
solar mass splitting and CP violation. In Section 3, we
demonstrate that the model, when evaluated at Point D, maintains
consistency with current experimental constraints and determine
the allowed parameter ranges. In Section 4, we discuss the
phenomenological implications of our results, focusing on the
Majorana nature of neutrinos, CP violation, and the potential for
leptogenesis. Finally, Section 5 provides a summary of our
findings and concluding remarks. Appendix A is devoted to a brief
introduction of the Twisted Friedberg Lee (TFL) symmetry.

%===============================================================================
%===============================================================================
\section{Theoretical Framework and Model Construction}\label{sec2}
%===============================================================================
%===============================================================================

In this section, we extend the standard Friedberg Lee (FL)
framework by introducing complex Yukawa coupling constants. This
extension induces CP violation through the generation of a
non-zero reactor angle, $U_{e3} \neq 0$. We impose the physical
requirement that the eigenvalues of the resulting generalized mass
matrix remain real. Under this constraint, the soft breaking of
$\mu-\tau$ symmetry is uniquely realized by the condition $a \in
\mathbb{R}$ and $b, c \in \mathbb{C}$ with $b = c^{\star}$. This
specific configuration leads to a non-Hermitian but symmetric mass
matrix, which preserves the reality of the mass spectrum. To
facilitate the subsequent analysis, we define the parameter
decomposition as:
\begin{equation}
    \text{Re}(b) = \text{Re}(c) = b_{r}, \quad \text{Im}(b) = -\text{Im}(c) = B,
\end{equation}
where $B$ serves as the fundamental parameter governing both the
degree of $\mu-\tau$ symmetry breaking and the magnitude of CP
violation\footnote{Given their role in these phenomena, we expect
$B$ to be a small parameter in the context of perturbative
analysis.}.

The generalized Majorana neutrino mass matrix, $M^{\prime}_{\nu}$,
is constructed as follows:
\begin{equation}\label{e8}
M^{\prime}_{\nu}=\left(\begin{array}{ccc}2b_{r}+m_{0} & -b_{r} & -b_{r}\\
-b_{r} & a+b_{r}+m_{0} & -a\\-b_{r}
& -a & a+b_{r}+m_{0}\end{array}\right)+iB\left(\begin{array}{ccc}0 & -1 & 1\\
-1 & 1 & 0\\1 & 0 & -1\end{array}\right).
\end{equation}
It is important to note that $M^{\prime}_{\nu}$ is a complex
symmetric matrix, which is the required form for a Majorana mass
term. Furthermore, both the generalized matrix $M^{\prime}_{\nu}$
and the original FL matrix $M_{FL}$ exhibit the magic property,
characterized by the existence of a common eigenvector
$(\frac{1}{\sqrt{3}}, \frac{1}{\sqrt{3}}, \frac{1}{\sqrt{3}})$. To
maintain consistency with the mixing structure required by
Eq.\,(\ref{etbm}), we identify this eigenvector with the second
neutrino mass eigenstate, $|\nu_{2}\rangle$.

To leading order in the parameter $B$, a direct diagonalization of
$M'_{\nu}$ yields the following approximate mass eigenvalues:
\begin{eqnarray} \vspace{.2cm}\label{e11}
\breve{m}_{1}&=&(a+2b_{r}+m_{0})+\sqrt{(a-b_{r})^{2}-3B^{2}}\nonumber\\\breve{m}_{2}&=&m_{0},
\nonumber\\\breve{m}_{3}&=&(a+2b_{r}+m_{0})-\sqrt{(a-b_{r})^{2}-3B^{2}}.
\end{eqnarray}
The expressions in Eq.~\eqref{e11} are valid in the limit $B \to
0$. By comparing these results with the constraints established in
Eq.~\eqref{e6}, we obtain the requirement that $a < b$ for
physical consistency.

Since $M'_{\nu}$ is a non-Hermitian symmetric matrix, its
diagonalization requires two distinct unitary matrices, $U$ and
$V$, such that $V = U^*$. These matrices can be formally obtained
by diagonalizing the Hermitian products $M'_{\nu}
M'^{\dagger}_{\nu}$ and $M'^{\dagger}_{\nu} M'_{\nu}$,
respectively.\footnote{In the context of Majorana neutrinos, $U$
and $V$ are related through the symmetry of the mass matrix,
ensuring the consistency of the Majorana nature.} The diagonalized
mass matrix, $M'_{\text{diag}} = U^{\dagger} M'_{\nu} V$, yields
the following complex mass eigenvalues:
\begin{eqnarray} \vspace{.2cm}\label{e111}
m'_{1}&=&\frac{iB(a-b_{r})+3B^{2}+(a+2b_{r}+m_{0})^{2}-(a-b_{r}+iB)
\sqrt{3B^{2}+(a+2b_{r}+m_{0})^{2}}}{a+2(b_{r}+iB)+m_{0}},
\nonumber\\m'_{2}&=&m_{0},\nonumber\\m'_{3}&=&\frac{iB(a-b_{r})+3B^{2}
+(a+2b_{r}+m_{0})^{2}+(a-b_{r}+iB)\sqrt{3B^{2}+(a+2b_{r}
+m_{0})^{2}}}{a+2(b_{r}+iB)+m_{0}},
\end{eqnarray}

Since the eigenvalues $m'_{1}$ and $m'_{3}$ are complex, we
parameterize the diagonal mass matrix by separating the physical
real masses from the Majorana CP violating phases. Following the
standard procedure for non-Hermitian Majorana matrices
\cite{phase1}, the most general form of the diagonalized mass
matrix $M'_{\text{diag}}$ can be expressed as:
\begin{equation}\label{ereal}
M^{\prime}_{diag}=e^{i\alpha}e^{i\beta\lambda_{3}}e^{i\gamma\lambda_{8}}M'^{real}_{diag},
\end{equation}
where $M_{\text{diag}}^{\text{real}} = \text{diag}(m_{1}, m_{2},
m_{3})$ contains the real, physical mass eigenvalues.

The overall phase $\alpha$ is unobservable and can be set to zero
without loss of generality; in our model, it vanishes naturally.
Furthermore, even if $\alpha$ were non-zero, it would not
influence the physical observables. Due to the fact that the
second mass eigenvalue $m'_{2}$ is strictly real, the Majorana
phases are constrained such that $\beta = \gamma$. This symmetry
implies that the phases of the remaining eigenvalues satisfy
$\arg(m'_{1}) = -\arg(m'_{3}) = 2\beta$\footnote{Note that
neglecting the overall phase is equivalent to requiring that
$\det(M'_{\text{diag}})$ is real and $\det(U)=1$ \cite{phase1}.}.
By imposing this phase condition on Eq.~\eqref{e11}, we obtain the
following constraint for the parameter $B$:
\begin{equation}\label{e12}
B=\pm\sqrt{\frac{-\left(2a+b_{r}+m_{0}\right)\left(3b_{r}+m_{0}\right)}{3}}.
\end{equation}

To ensure that the parameter $B$ remains real, the parameter
$b_{r}$ must satisfy the following inequality:
\begin{equation}
-\frac{m_{0}}{3} \leq b_{r} \leq -(2a + m_{0}).
\end{equation}
The lower bound of this interval is consistent with the
requirement $m_{1} > 0$ as specified in Eq.~\eqref{e6}.
Furthermore, since the unitary matrices $U$ and $V$ must recover
the Tri-Bimaximal Mixing (TBM) structure in the limit $B \to 0$
(see Eq.~\eqref{etbm}), the condition $2b_{r} + a + m_{0} \geq 0$
is imposed. Combining this with the constraint $a < b$ from
Eq.~\eqref{e12}, we arrive at the requirement $3a + m_{0} \leq 0$.

\begin{center}
\begin{figure}[th] \includegraphics[width=6cm]{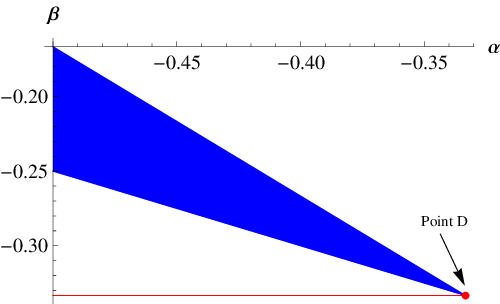}
\caption{\label{fig.1} \small
   The allowed parameter space in the
$\alpha, \beta$ plane, where $\alpha \equiv a/m_{0}$ and $\beta
\equiv b_{r}/m_{0}$. The shaded triangular region denotes the
domain where CP violation occurs, which is bounded by the line
$2\beta + \alpha + 1 = 0$. The red dot represents Point D at
$(-\frac{1}{3}, -\frac{1}{3})$, which corresponds to the CP
conserving limit where $B=0$.}
  \label{geometry}
\end{figure}
\end{center}

In light of the allowed parameter regions and the inherent
symmetry of the Friedberg Lee (FL) model, it is found that $b_{r}
< 0$\footnote{This result is in excellent agreement with solar
neutrino oscillation data, which necessitates the mass ordering
$m_{2} > m_{1}$.}. It should be noted that CP violation is not
present across the entire parameter space; instead, it is confined
to a specific region within the $a, b_{r}$ plane for $B \neq 0$.
The allowed region for the onset of CP violation is illustrated in
Fig.~\ref{fig.1}.

%\begin{figure}[th]
%\centering
%\includegraphics[width=6cm]{cp22a}
%\caption{\label{fig.1} \small The allowed parameter space in the
%$\alpha, \beta$ plane, where $\alpha \equiv a/m_{0}$ and $\beta
%\equiv b_{r}/m_{0}$. The shaded triangular region denotes the
%domain where CP violation occurs, which is bounded by the line
%$2\beta + \alpha + 1 = 0$. The red dot represents Point D at
%$(-\frac{1}{3}, -\frac{1}{3})$, which corresponds to the CP
%conserving limit where $B=0$.} \label{geometry}
%\end{figure}

%\begin{center}
%\begin{figure}[th] \includegraphics[width=6cm]{CP22a}
%\caption{\label{fig:1} \small
 % The allowed parameter space in the
%$\alpha, \beta$ plane, where $\alpha \equiv a/m_{0}$ and $\beta
%\equiv b_{r}/m_{0}$. The shaded triangular region denotes the
%domain where CP violation occurs, which is bounded by the line
%$2\beta + \alpha + 1 = 0$. The red dot represents Point D at
%$(-\frac{1}{3}, -\frac{1}{3})$, which corresponds to the CP
%conserving limit where $B=0$. }

%  \label{geometry}
%\end{figure}
%\end{center}

A comprehensive analysis of this parameter space, encompassing the
resulting CP violating effects, their phenomenological
implications, and their consistency with current experimental
constraints, has been detailed in Ref.~\cite{me1}.

A significant feature in the parameter space is highlighted by the
red marker in Fig.~\ref{fig.1}, which corresponds to the limit $a
= b_{r} = -m_{0}/3$. In this limit, the perturbation parameter $B$
vanishes, thereby effectively restoring the CP conserving TBM
limit. This transition point is hereafter referred to as Point D.

The location of Point D is of particular physical interest, as it
represents the limit where the mass squared difference vanishes,
i.e., $\Delta m^2 = 0$. This singularity marks the transition to a
degenerate mass spectrum. In this section, we focus our analysis
on Point D (see Fig.~\ref{fig.1}), where the generalized Friedberg
Lee neutrino mass matrix $M'_{\nu}$ in Eq.~(\ref{e8}) reduces to:
\begin{equation}\label{e22}
{M_{\nu}}|_{\text{point D}} = (3b_{r}+m_{0})\mathbf{1} -
b_{r}\mathbf{D},
\end{equation}
where $\mathbf{1}$ denotes the identity matrix and $\mathbf{D}$ is
the democratic matrix with all entries equal to unity. At this
critical point, for the specific case $3b_{r} + m_{0} = 0$, the
first term vanishes, reducing the mass matrix to a purely
democratic form, $M_{\nu} = -b_{r}\mathbf{D}$. Consequently,
$M_{\nu}$ adopts a democratic structure, which is naturally
realized by the $S_{3}$ permutation symmetry \cite{demo1}. By
independently imposing $S_{3}$ as a flavor symmetry for both the
left handed and right handed neutrino sectors, denoted by $S_{3L}$
and $S_{3R}$, the mass matrix in Eq.~(\ref{e22}) becomes invariant
under the product group $S_{3L} \times S_{3R}$.

As shown in Eq.~(\ref{e22}), the mass matrix $M_{\nu}$ is
diagonalized by the $U_{\text{TBM}}$ matrix, yielding the mass
eigenvalues $m_2=m_0$ and $m_1=m_3=0$. Given that this mass
spectrum is phenomenologically disfavored by current experimental
data (which require non-vanishing mass splittings), it is
necessary to modify the matrix to allow for $m_1, m_3 \neq 0$. To
this end, we introduce a symmetry breaking term into the FL
framework.

The integration of FL symmetry with $\mu-\tau$ symmetry yields a
well established class of translational symmetry known as the TFL
symmetry \cite{TFL}. For Dirac neutrinos, we separately impose
this TFL symmetry on the left handed and right handed sectors:
\begin{eqnarray}\label{eq:RFL}
\nu_{Li} \rightarrow \nu_{Li}^{'}=S_{ij}^L\nu_{Lj}+\Lambda_{Lj} z,
\quad \nu_{Ri} \rightarrow
\nu_{Ri}^{'}=S_{ij}^R\nu_{Rj}+\Lambda_{Rj} z,
\end{eqnarray}
where $\Lambda = (\Lambda_1, \Lambda_2, \Lambda_3)^T$ represents a
vector of c numbers, and $z$ is a space time independent Grassmann
parameter satisfying $z^2=0$. The matrix $S$ denotes the
permutation between the second and third generations:
\begin{equation}
 S=\left(\begin{array}{ccc} 1 & 0 & 0 \\ 0 & 0 & 1 \\ 0 & 1 & 0 \end{array}\right).
\end{equation}

As detailed in Appendix~\ref{app:TFL_symmetry}, the imposition of
these transformations upon the $S_{3L} \times S_{3R}$ flavor
symmetry of the mass matrix $M_{\nu}$ induces a symmetry breaking
mass matrix, $M^B_{\nu}$:
\begin{equation}
 M^B_{\nu}=g\left(\begin{array}{ccc} 4 & -2 & -2 \\ -2 & 1 & 1 \\ -2 & 1 & 1 \end{array}\right). \label{eq:eFL4}
\end{equation}

By incorporating the symmetry breaking term $M^B_\nu$ from
Eq.~(\ref{eq:eFL4}) into the democratic mass matrix $M_\nu$ in
Eq.~(\ref{e22}), the total neutrino mass matrix at Point D is
expressed as:
\begin{equation}
 {M_\nu^{BD}}|_{\text{point D}}=\frac{m_0}{3} \left(\begin{array}{ccc} 1 & 1 & 1 \\ 1 & 1 & 1 \\ 1 & 1 & 1 \end{array}\right)
 + g \left(\begin{array}{ccc} 4 & -2 & -2 \\ -2 & 1 & 1 \\ -2 & 1 & 1 \end{array}\right).\label{eq:MD}
\end{equation}
This configuration, which we denote as the Broken Democratic (BD)
neutrino mass matrix, yields the following mass eigenvalues:
\begin{equation}
\tilde{m}_1 = 6g, \quad \tilde{m}_2 = m_0, \quad \tilde{m}_3 =
0.\label{eq:MD1}
\end{equation}

In summary, the proposed model exhibits several key features: (i)
the TFL symmetry serves as a natural mechanism for breaking the
$S_{3L} \times S_{3R}$ flavor symmetry; (ii) while the total mass
matrix in Eq.~(\ref{eq:MD}) breaks the $S_{3L} \times S_{3R}$
symmetry, it simultaneously preserves both the $\mu-\tau$
reflection symmetry and the magic property, thereby ensuring that
$U_{\text{TBM}}$ is realized as the mixing matrix; and (iii) the
resulting neutrino mass spectrum is characterized by an inverted
hierarchy.

To move beyond the idealized TBM limit and obtain a more realistic
neutrino mixing pattern, our primary objective is to introduce a
small perturbative mass matrix to the Broken Democratic (BD)
matrix, $M_\nu^{BD}$, given in Eq.~(\ref{eq:MD}). To facilitate
this systematic expansion, we first consider the most general
structure of a mass matrix $\mathcal{M}$ that is diagonalized by
$U_{\text{TBM}}$ in the flavor basis:
\begin{equation}\label{egf0}
\mathcal{M} = U_{\text{TBM}} \begin{pmatrix} m_1 & 0 & 0 \\ 0 &
m_2 & 0 \\ 0 & 0 & m_3 \end{pmatrix} U_{\text{TBM}}^T =
\begin{pmatrix} A & B & B \\ B & A+C & B-C \\ B & B-C & A+C
\end{pmatrix},
\end{equation}
where the coefficients $A, B,$ and $C$ are related to the mass
eigenvalues via:
\begin{equation}\label{eABC}
A = m - \frac{\Delta_{31}}{3}, \quad B = \frac{\Delta_{31} -
\Delta_{32}}{3}, \quad \text{and} \quad C = \frac{\Delta_{31}}{2}.
\end{equation}
Here, $m$ represents the average mass, and $\Delta_{3j} \equiv m_3
- m_j$ denotes the mass splittings for $j=1,2$. The average mass
$m$ and the splittings $\Delta_{3j}$ are defined as:
\begin{equation}\label{egf}
m = \frac{\sum_{i=1}^3 m_i}{3}, \quad \Delta_{3j} = m_3 - m_j.
\end{equation}

By substituting the eigenvalues $\tilde{m}_i$ obtained from the BD
matrix in Eq.~(\ref{eq:MD1}) into these expressions, we find:
\begin{equation}
m = \frac{m_0}{3} + 2g, \quad \Delta_{32} = -m_0, \quad \text{and}
\quad \Delta_{31} = -6g.
\end{equation}
We work in the flavor basis, where the lepton mixing is entirely
determined by the structure of the neutrino mass matrix.

It is well established from experimental observations that the
solar mass squared difference, $\Delta m^2_{21} = m^2_2 - m^2_1$,
is positive and significantly smaller than the atmospheric scale.
Applying the eigenvalues $\tilde{m}_i$ from Eq.~(\ref{eq:MD1}) to
the expression for $\Delta m^2_{21}$ yields the phenomenological
constraint $|g| < m_0/6$.

In the regime where the atmospheric mass splittings are
approximately degenerate, i.e., $\Delta_{32} \simeq \Delta_{31}
\equiv \Delta$, our relations imply $\Delta \simeq -m_0 \simeq
-6g$. This condition leads to $g \simeq m_0/6$ with $\Delta < 0$,
a result that is entirely consistent with an inverted neutrino
mass ordering. In this context, $\Delta$ plays a fundamental role
by setting the scale for atmospheric neutrino oscillations.

Under these approximations, the Broken Democratic (BD) neutrino
mass matrix $M_\nu^{BD}$ at Point D can be expressed in the flavor
basis as:
\begin{equation}\label{eunp}
{M^{0}_{\nu}}|_{\text{point D}} \simeq \begin{pmatrix} m -
\frac{\Delta}{3} & 0 & 0 \\ 0 & m + \frac{\Delta}{6} &
-\frac{\Delta}{2} \\ 0 & -\frac{\Delta}{2} & m + \frac{\Delta}{6}
\end{pmatrix},
\end{equation}
where $M_\nu^{(0)}$ denotes the unperturbed neutrino mass matrix
at Point D. Since this unperturbed matrix preserves both the magic
and $\mu, \tau$ symmetries, the resulting mixing matrix remains
exactly $U_{\text{TBM}}$.

The eigenvalues of the unperturbed mass matrix $M_\nu^{(0)}$ in
Eq.~(\ref{eunp}) are given by:
\begin{equation}\label{emm2}
m^{(0)}_1 \simeq m^{(0)}_2 = m-\frac{\Delta}{3}=m_0, \quad
\text{and} \quad m^{(0)}_{3}=m+\frac{2\Delta}{3}=0.
\end{equation}
While $m^{(0)}_1$ and $m^{(0)}_2$ are real and positive, the solar
mass splitting vanishes at this level of approximation.
Furthermore, since $m^{(0)}_3=0$, the resulting mass spectrum
corresponds to an inverted ordering, which is consistent with the
results presented in Ref.~\cite{me1}.

Although this stage provides a reasonable first approximation to
the mixing pattern via the TBM limit, the model exhibits
significant phenomenological shortcomings. First, the solar mass
squared difference $\Delta m^2_{21}$ is absent. Second, the mixing
matrix remains strictly tribimaximal ($U_{\mathrm{TBM}}$). Third,
the mass spectrum contains a degeneracy between $m_1$ and $m_2$ in
the limit where $m_1=m_2=m_0$, which deviates from both the
generalized FL model discussed in Ref.~\cite{me1} and current
experimental observations \cite{exp} that require distinct mass
values for the individual states.

A crucial question arises regarding the breaking of the $\mu,
\tau$ symmetry, which is a prerequisite for CP violation within
this framework. Specifically, we investigate whether the
atmospheric mixing angle $\theta_{23}$ remains maximal at
$45^\circ$ following this symmetry breaking.

Our objective in the subsequent sections is to resolve these
issues by introducing a small perturbative mass matrix to the
Broken Democratic (BD) matrix $M_\nu^{BD}$ in Eq.~(\ref{eq:MD}).
This perturbation is specifically designed to achieve several
goals. It is intended to generate a non-vanishing solar mass
squared difference $\Delta m^2_{21}$, induce a non-zero $U_{e3}$
entry corresponding to $\theta_{13} \neq 0$, and generate a
leptonic CP violating phase $\delta$. Additionally, the
perturbation aims to provide small corrections to $\theta_{12}$
while leaving $\theta_{23}$ essentially unchanged.

Given that both $\theta_{13}$ and $\Delta m^2_{21}$ are small in
comparison to the atmospheric scale, this perturbative approach
offers a systematic and controlled method to derive realistic
corrections to the TBM pattern. To the best of our knowledge, the
exploration of such a generalized FL model at this specific point
in the parameter space has not been previously addressed in the
literature.

Our approach to constructing the neutrino mass matrix at Point D
is rooted in perturbation theory. We decompose the total mass
matrix as:
\begin{equation}
\left. M_\nu \right|_{\text{point D}} = M_\nu^{0} + M_\nu^{P},
\end{equation}
where $M_\nu^{P}$ represents a perturbation matrix satisfying the
hierarchy $\|M_\nu^{P}\| \ll \|M_\nu^{0}\|$. While the unperturbed
matrix $M_\nu^{0}$ in Eq.~(\ref{eunp}) is real and symmetric,
which implies it is also Hermitian, we consider a general complex
perturbation matrix $M_\nu^{P}$. Such a complex structure is
essential to induce a non-zero $\theta_{13}$, a non-trivial
leptonic CP violating phase $\delta$, and a non-vanishing solar
mass splitting.

In the unperturbed limit, the mass eigenstates in the mass basis
are given by:
\begin{equation}
\label{em0} |\nu^{(0)}_{1}\rangle=
\begin{pmatrix}
1\\
0\\
0
\end{pmatrix},
\qquad |\nu^{(0)}_{2}\rangle=
\begin{pmatrix}
0\\
1\\
0
\end{pmatrix},
\qquad |\nu^{(0)}_{3}\rangle=
\begin{pmatrix}
0\\
0\\
1
\end{pmatrix}.
\end{equation}
At this level, the states $|\nu^{(0)}_{1}\rangle$ and
$|\nu^{(0)}_{2}\rangle$ are degenerate. To lift this $m_1, m_2$
degeneracy and generate the required solar scale, we impose the
condition:
\begin{equation}
\langle\nu_{i}^{(0)}|M_{\nu}^{P}|\nu_{j}^{(0)}\rangle =
m_{i}^{(1)}\delta_{ij}, \quad (i,j=1,2),
\end{equation}
where $m_1^{(1)} \neq m_2^{(1)}$ ensures a non-vanishing solar
mass splitting. To maintain the perturbative hierarchy and focus
on the essential degrees of freedom, we set $(M_\nu^P)_{33} = 0$
and concentrate on the elements $(M_\nu^P)_{13}$ and
$(M_\nu^P)_{23}$. This strategic choice is motivated by the
necessity to simultaneously generate a non-zero $\theta_{13}$ and
provide the required corrections to accommodate the solar mixing
angle.

In the flavor basis, the unperturbed eigenstates
$|\nu^{(0)}_{i}\rangle$ correspond to the columns of the TBM
mixing matrix defined in Eq.~(\ref{etbm}). Consequently, these
states are expressed as:
\begin{equation}
\label{em00} |\nu^{(0)}_{1}\rangle=
\begin{pmatrix}
\sqrt{2/3} \\
-1/\sqrt{6} \\
1/\sqrt{6}
\end{pmatrix}, \quad
|\nu^{(0)}_{2}\rangle=
\begin{pmatrix}
1/\sqrt{3} \\
1/\sqrt{3} \\
-1/\sqrt{3}
\end{pmatrix}, \quad
|\nu^{(0)}_{3}\rangle=
\begin{pmatrix}
0 \\
1/\sqrt{2} \\
1/\sqrt{2}
\end{pmatrix}.
\end{equation}

Our primary objective is to ensure that the third perturbed mass
eigenstate, in the presence of CP violation, reproduces the third
column of the experimental mixing matrix as defined in
Eq.~(\ref{emixing}) when projected onto the flavor basis. To
achieve this, rather than postulating an arbitrary form for the
perturbation, we adopt an inversion procedure. In this method, we
reconstruct the elements of the perturbation matrix $M_\nu^P$ by
requiring that it yields the desired physical eigenstates and the
observed mass spectrum.

We assume that the perturbation matrix $M_\nu^{P}$ is complex and
symmetric, which implies that it is non-Hermitian. Consequently,
the total mass matrix $M_\nu = M_\nu^{0} + M_\nu^{P}$ also becomes
non-Hermitian. This non-Hermiticity is a crucial feature of our
model because it provides the necessary degrees of freedom to
generate both a non-zero $\theta_{13}$ and a non-trivial leptonic
CP violating phase $\delta$.

To determine the perturbed eigenstates, we analyze the Hermitian
product $M_\nu^\dagger M_\nu$. By expanding this product and
neglecting terms of order $\mathcal{O}(|M_\nu^P|^2)$, we obtain
the following approximation:
\begin{equation}
M_\nu^\dagger M_\nu \approx M_\nu^{0\dagger}M_\nu^0 +
M_\nu^{0\dagger}M_\nu^P + M_\nu^{P^{\dagger}}M_\nu^0.
\end{equation}
The unperturbed term $M_\nu^{0\dagger}M_\nu^0$ is Hermitian,
possessing eigenvalues $|m_{i}^{(0)}|^{2}$ and eigenstates that
correspond to the columns of the TBM mixing matrix
$U_{\text{TBM}}$ defined in Eq.~(\ref{etbm}). Within the
perturbative framework, we retain terms up to linear order in
$s_{13}$. Consequently, the third perturbed mass eigenstate can be
expressed as:
\begin{equation}
|\nu_3\rangle=|\nu^{(0)}_3\rangle+\sum_{j \neq 3} {\cal
C}_{3j}|\nu^{(0)}_j\rangle, \label{eper1}
\end{equation}
where the complex coefficients ${\cal C}_{3j}$ are given by:
\begin{equation}
{\cal C}_{3j}= -{\cal C}^*_{j3}=\left(|m^{(0)}_3|^2 -
|m^{(0)}_j|^2\right)^{-1}{\cal M}_{j3}, \quad (j \neq 3),
\label{eper3}
\end{equation}
with the transition matrix elements defined as ${\cal M}_{j3}={
\langle\nu^{(0)}_j| ( M_\nu^{0\dagger}M_\nu^P +
M_\nu^{P^{\dagger}}M_\nu^0) |\nu^{(0)}_3\rangle }$. These
coefficients ${\cal C}_{3j}$ are inherently complex and are
proportional to ${\cal M}_{j3}$ in the mass basis \cite{meuni}.

The state $|\nu_3\rangle$ in Eq.~(\ref{eper1}) must coincide with
the third column of the experimental neutrino mixing matrix $U$ as
defined in Eq.~(\ref{emixing}). By expanding Eq.~(\ref{eper1}) in
the flavor basis and comparing it with the target mixing pattern,
we arrive at the following matrix equation:
\begin{equation}\label{e1}
\left(\begin{array}{c}s_{13}e^{-i\delta}\\
s_{23}c_{13}\\c_{23}c_{13}
\end{array}\right)=\left(\begin{array}{c}0\\
-\frac{1}{\sqrt{2}}\\\frac{1}{\sqrt{2}}\end{array}\right)+\left(\begin{array}{c}\frac{-\sqrt{2}{\cal C}_{31}+{\cal C}_{32}}{\sqrt{3}}\\
\frac{{\cal C}_{31}}{\sqrt{6}}+\frac{{\cal
C}_{32}}{\sqrt{3}}\\\frac{{\cal C}_{31}}{\sqrt{6}}+\frac{{\cal
C}_{32}}{\sqrt{3}}\end{array}\right).
\end{equation}
In the limit of maximal $\mu, \tau$ symmetry where
$\theta_{23}=45^\circ$, and to linear order in $s_{13}$, the
coefficients are determined to be:
\begin{equation}
{\cal C}_{31}=-\sqrt{\frac{2}{3}} s_{13}e^{-i\delta}, \quad
\text{and} \quad {\cal C}_{32}=\sqrt{\frac{1}{3}}
s_{13}e^{-i\delta}.
\end{equation}
Finally, by applying Eqs.~(\ref{emm2}) and (\ref{eper3}) within
the mass basis, we extract the required components of the
perturbation matrix:
\begin{equation}
{(M^{P}_\nu)}_{13}= m_0\sqrt{\frac{2}{3}} s_{13}e^{-i\delta},
\quad \text{and} \quad {(M^{P}_\nu)}_{23}= -m_0\sqrt{\frac{1}{3}}
s_{13}e^{-i\delta}.
\end{equation}

Having successfully generated a non-zero $\theta_{13}$ via the
aforementioned perturbative approach, we now address the solar
mass splitting. Within our framework of a minimal perturbation
matrix, we impose the constraint $(M_\nu^P)_{12} = (M_\nu^P)_{21}
= 0$. The first order corrections to the neutrino masses are
governed by the relation $m^{(1)}_i\delta_{ij} =
\langle\nu^{(0)}_i|M_{\nu}^P|\nu^{(0)}_j\rangle$. To lift the
solar degeneracy while maintaining the stability of the other mass
scales, we require:
\begin{equation}
m^{(1)}_1 = m^{(1)}_3 = 0 \quad \text{and} \quad m^{(1)}_2 \neq 0.
\label{efirstm}
\end{equation}
In the mass basis, this condition necessitates a non-vanishing
$(M_\nu^P)_{22}$ component, while the remaining diagonal elements
of the perturbation matrix must vanish. This specific correction
is responsible for lifting the $m_1, m_2$ degeneracy, such that
$m^{(1)}_2 = m_2 - m_1$, which subsequently ensures a positive
solar mass splitting $\Delta m^2_{21} > 0$.

Consequently, the complete form of the perturbation mass matrix at
Point D in the mass basis is expressed as:
\begin{equation}\label{eM2}
{M_\nu^P}|_{\text{point D}} = m_0 s_{13}
\begin{pmatrix}
0 & 0 & \sqrt{\frac{2}{3}}e^{-i\delta} \\
0 & \varepsilon & -\sqrt{\frac{1}{3}}e^{-i\delta} \\
\sqrt{\frac{2}{3}}e^{-i\delta} & -\sqrt{\frac{1}{3}}e^{-i\delta} &
0
\end{pmatrix},
\end{equation}
where $\varepsilon \equiv m^{(1)}_2 / (m_0 s_{13})$ is a
dimensionless parameter characterizing the relative scale between
the solar mass splitting and the reactor mixing angle. This
parameter $\varepsilon$ will be utilized in the subsequent section
to estimate the magnitude of $s_{13}$.

In a general setting, the solar mass correction $m^{(1)}_2$ may be
complex, which we parametrize as $m^{(1)}_2 \equiv |m^{(1)}_2|
\exp(i \varphi)$. Accordingly, the physical mass $m_2$ is
expressed as:
\begin{equation}
m_2 = m^{(0)}_2 + m^{(1)}_2 \equiv |m_2| \exp(i \phi),
\end{equation}
where $\phi$ denotes the phase of the second mass eigenvalue. At
Point D, the neutrino mass spectrum is determined by:
\begin{align}
|m_1| &= m_0, \label{em2_1} \\
|m_3| &= 0, \label{em2_2} \\
|m_2| &= \sqrt{(m^{(0)}_1)^2 + |m^{(1)}_2|^2 + 2 m^{(0)}_1 |m^{(1)}_2| \cos\varphi}, \label{em2_3} \\
\phi &= \tan^{-1} \left[ \frac{|m^{(1)}_2| \sin\varphi}{m^{(0)}_1
+ |m^{(1)}_2| \cos\varphi} \right]. \label{em2_4}
\end{align}
While the phase $\phi$ can be rephased away in the case of Dirac
neutrinos, it remains a physical degree of freedom for Majorana
neutrinos, manifesting as a Majorana phase that contributes to CP
violation.

We now perform a basis transformation to express the perturbation
mass matrix $M_\nu^P$ in the flavor basis. Utilizing the standard
transformation $M_\nu^{P^{(f)}} = U_{\text{TBM}} M_\nu^P
U_{\text{TBM}}^T$, the flavor basis representation of the
perturbation at Point D is obtained as:
\begin{equation}\label{eM3}
{M_\nu^{P^{(f)}}}|_{\text{point D}} = m_0 s_{13} \left[
\frac{e^{-i\delta}}{\sqrt{2}}
\begin{pmatrix}
0 & 1 & -1 \\
1 & 0 & 0 \\
-1 & 0 & 0
\end{pmatrix} + \frac{\varepsilon}{3}
\begin{pmatrix}
1 & 1 & 1 \\
1 & 1 & 1 \\
1 & 1 & 1
\end{pmatrix} \right],
\end{equation}
where the first term is responsible for generating the reactor
mixing angle $\theta_{13}$, while the second term accounts for the
solar mass splitting. It is noteworthy that the structure of
$M_\nu^{P^{(f)}}$ in Eq.~(\ref{eM3}) explicitly breaks both the
magic and $\mu, \tau$ symmetries inherent in the unperturbed mass
matrix.

To derive the neutrino mixing matrix at Point D, we apply
degenerate perturbation theory \cite{Schiff} to the mass matrix
$M_\nu^P$ given in Eq.~(\ref{eM2}), working to first order in
$s_{13}$. The resulting mixing matrix, which incorporates the
leptonic CP violating phase, is given by:
\begin{equation}\label{eU}
{U}|_{\text{point D}} = U_{\text{TBM}} + s_{13}e^{i\delta}
\begin{pmatrix}
0 & 0 & e^{-2i\delta} \\
-\sqrt{1/3} & \sqrt{1/6} & 0 \\
\sqrt{1/3} & -\sqrt{1/6} & 0
\end{pmatrix},
\end{equation}
where the unitarity of $U$ is preserved up to
$\mathcal{O}(s_{13})$. An analogous mixing structure has been
explored in Ref.~\cite{18} via an alternative approach, where the
compatibility of the model with experimental mixing data was also
demonstrated.

A rephasing invariant measure of CP violation in neutrino
oscillations is provided by the Jarlskog invariant $J$ \cite{J},
which is independent of the Dirac or Majorana nature of the
neutrinos. By substituting Eq.~(\ref{eU}) into the definition of
$J$ in Eq.~(\ref{eJ}), the expression at Point D simplifies to:
\begin{equation}\label{eJJ}
  J|_{\text{point D}} = -\frac{1}{3\sqrt{2}} s_{13} \sin \delta.
\end{equation}
This result demonstrates that a non-vanishing CP violation in the
lepton sector is realized provided that both $s_{13}$ and $\delta$
are non-zero.

%=============================================================
%=============================================================
\section{Numerical Results and Phenomenological Analysis}\label{sec3}
%=============================================================

In this section, we compare the results obtained from our
generalized FL model at the remarkable Point D, where $B=0$, with
the available experimental data presented in Eq.~(\ref{exp}). It
is worth noting that, in the generalized FL model \cite{me1},
various numerical predictions for the neutrino parameters arise
when $B \neq 0$. Therefore, we perform a systematic comparison of
our present results with those reported in \cite{me1} in two
distinct steps.

First, we determine the allowed ranges of the parameters of the
neutrino mass matrix, including the perturbation term, by
comparing our neutrino mass results with the constraints on
$\delta m^{2}$ and $\Delta m^{2}$ derived from the experimental
data \cite{exp}. The allowed ranges for $m_0$ and $g$ are obtained
as
\begin{eqnarray}\label{em0g}
m_0 &\approx& \pm(4.87-5.03) \times 10^{-2}\,\text{eV}, \nonumber\\
g &\approx& \pm(0.812-0.838) \times 10^{-2}\,\text{eV},
\end{eqnarray}
which are in excellent agreement with the ranges of $m_1$ and
$m_0$ reported in \cite{me1}. Thus, both the $m_0 > 0$ and $m_0 <
0$ branches remain permissible.

Figures~\ref{fig.2} and \ref{fig.3} illustrate the constraints
imposed by $\delta m^{2}$ on the $m^{(1)}_2 - \varphi$ parameter
space for the $m_0 > 0$ and $m_0 < 0$ branches, respectively. By
plotting the overlap between the experimental values of $\delta
m^{2}$ from Eq.~(\ref{exp}) and the predictions of our model from
Eq.~(\ref{em2_1}) and Eq.~(\ref{em2_2}), we initially obtain the
following broad range:
\begin{equation}\label{empp}
m^{(1)}_2 \approx \pm (0.073-0.9) \times 10^{-2}\,\text{eV}.
\end{equation}

Figure~\ref{fig.2} illustrates the limits imposed by $\delta
m^{2}$ and the condition $m_0>0$ on $m^{(1)}_2$, whereas
Figure~\ref{fig.3} shows the corresponding constraints for the
$m_0<0$ case. In both figures, we plot the overlap between the
experimental values of $\delta m^{2}$ from Eq.~(\ref{em2}) and the
allowed ranges of $m_0$ from Eq.~(\ref{em0g}) within the
$m^{(1)}_2-\varphi$ perturbation parameter space. For $m_0>0$, the
resulting range of the perturbation parameter is
\begin{equation}\label{empp}
m^{(1)}_2\approx (0.073-0.9)10^{-2}\text{eV},
\end{equation}
whereas, for $m_0<0$, the result is given by
\begin{equation}\label{empq}
m^{(1)}_2\approx -(0.073-0.9)10^{-2}\text{eV}.
\end{equation}

\begin{figure}[th]
\centering
\includegraphics[width=8cm]{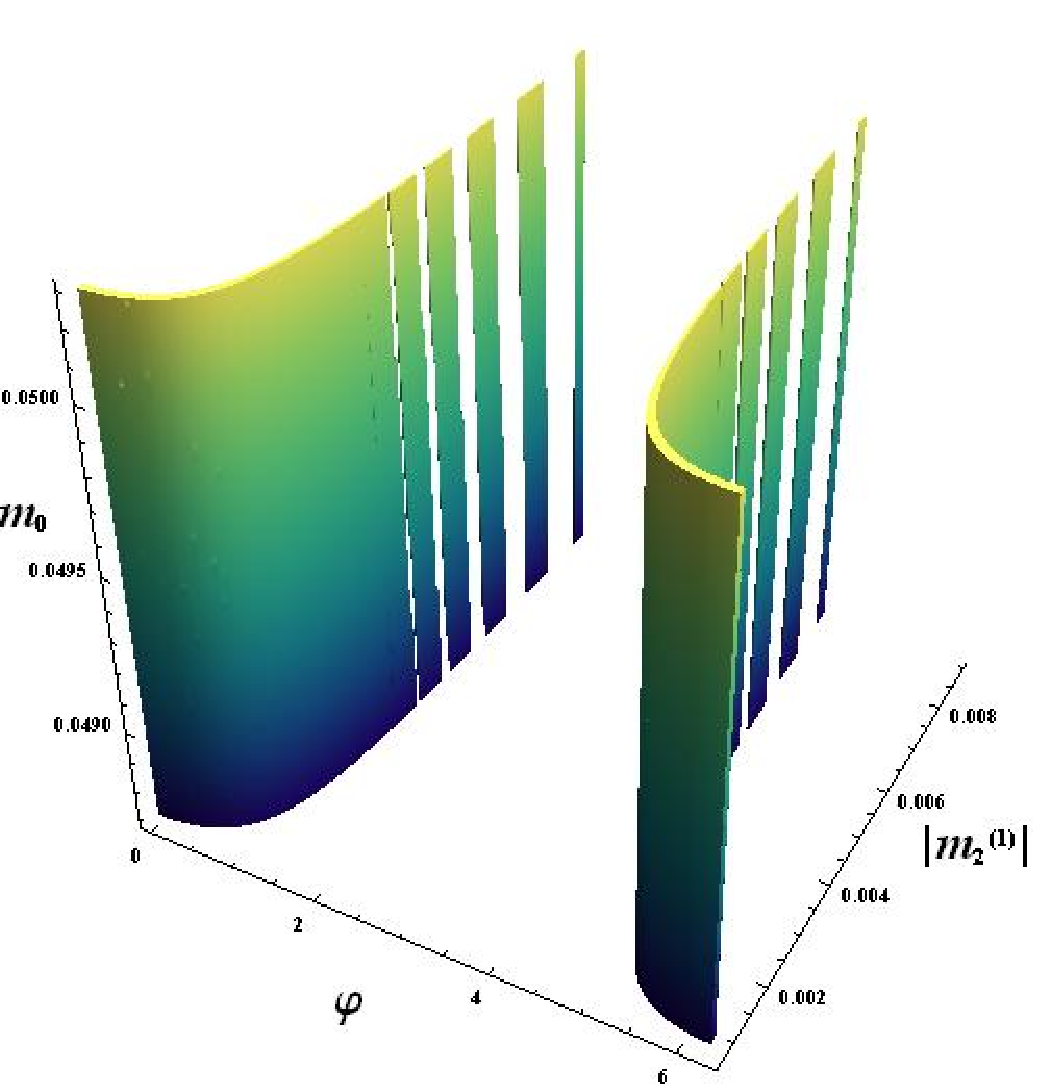}
\caption{\label{fig.2} \small (color online). The allowed
$m^{(1)}_2$--$\varphi$ parameter space for the $m_0 > 0$ branch.
The colored curves represent constant values of the mass scale
within the range $m_0 \in (4.87 \text{--} 5.03) \times
10^{-2}$~eV. The intersections of the model predictions with the
experimental $\Delta m^2_{21}$ constraints result in two distinct,
approximately symmetric compact regions. This symmetry suggests a
dual solution regime for the solar mass scale in the positive
$m_0$ branch.} \label{fig.2}
\end{figure}

\begin{figure}[th]
\centering
\includegraphics[width=8cm]{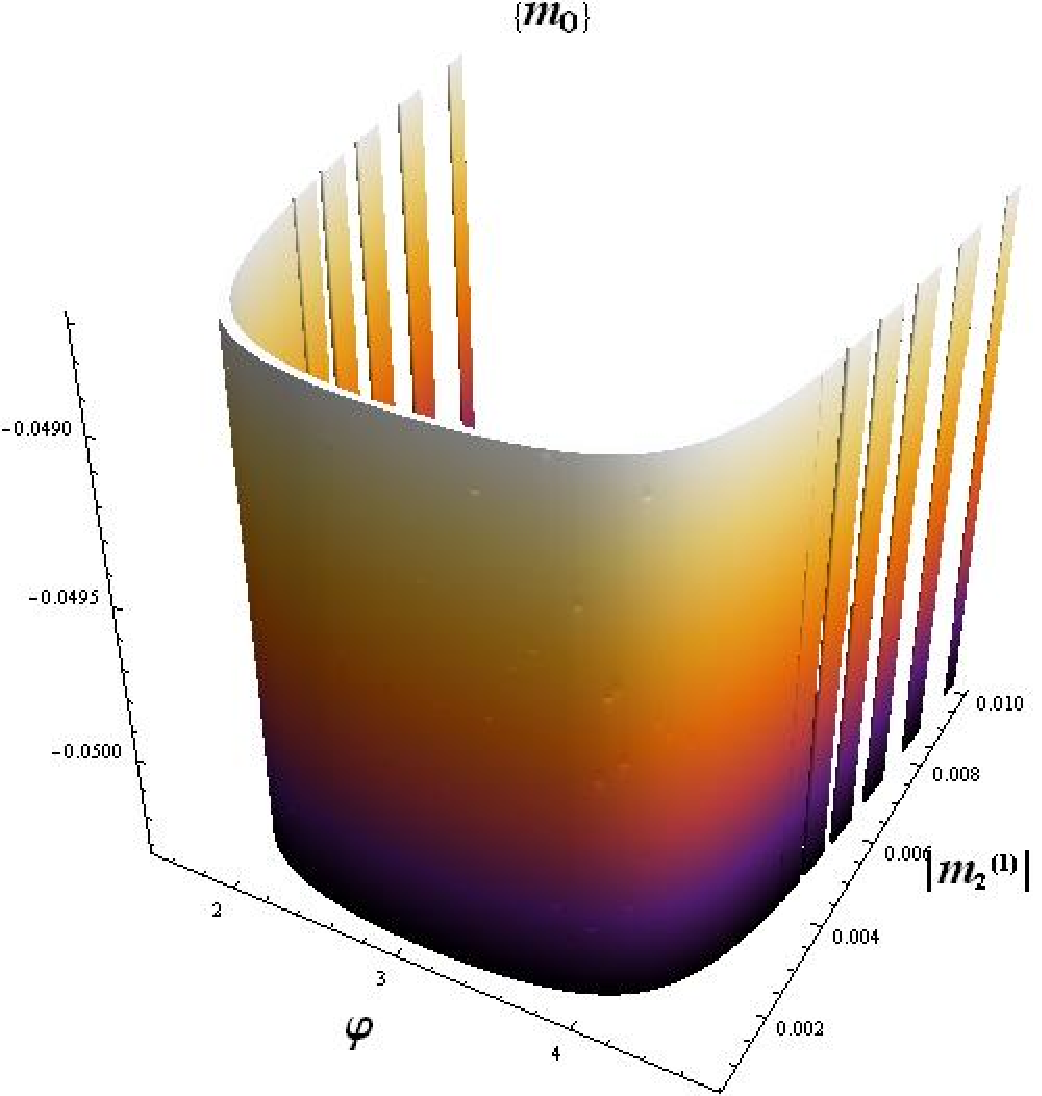}
\caption{\label{fig.3} \small (color online). The allowed
$m^{(1)}_2$--$\varphi$ parameter space for the $m_0 < 0$ branch.
The curves correspond to the range $m_0 \in -(4.87 \text{--} 5.03)
\times 10^{-2}$~eV. Unlike the $m_0 > 0$ case, the overlap with
the experimental $\Delta m^2_{21}$ values is restricted to a
single, highly localized, tiny region, indicating a substantially
more constrained parameter space.} \label{fig.3}
\end{figure}

A crucial observation is that the values of $m^{(1)}_2$ must also
reproduce the measured $\sin \theta_{13}$ within the experimental
bounds. We define the parameter $\varepsilon \equiv m^{(1)}_2 /
(m_0 s_{13})$ and, assuming that all nonzero components of the
perturbation matrix in Eq.~(\ref{eM2}) are of a similar order, we
expect $\varepsilon \sim \mathcal{O}(1)$. Given the established
scales of $m_0$ and $m^{(1)}_2$, this consistency check requires
$\sin \theta_{13} \sim \mathcal{O}(10^{-1})$.

Upon closer examination, we find that the entire range of
$m^{(1)}_2$ presented in Eq.~(\ref{empp}) is not fully compatible
with the experimentally measured order of magnitude of $\sin
\theta_{13}$. By imposing this stricter consistency requirement,
the valid ranges of $m^{(1)}_2$ and $\varphi$ are significantly
narrowed to
\begin{eqnarray}\label{empp1}
|m^{(1)}_2| &\approx& (0.487-0.9) \times 10^{-2}\,\text{eV}, \nonumber \\
\varphi & \approx& \begin{cases}
(83.12^{\circ}-91.72^{\circ}) \text{ or } (269.43^{\circ}-275.16^{\circ}), & \text{for } m_0 > 0, \\
(88.85^{\circ}-97.45^{\circ}) \text{ or }
(263.69^{\circ}-269.43^{\circ}), & \text{for } m_0 < 0.
\end{cases}
\end{eqnarray}

%^^^^^^
%=============================================================
\subsection{Summary of Results}
%=============================================================
In conclusion, we have successfully determined the complete set of
model parameters at the remarkable Point D, where $B=0$. The
interplay between the unperturbed mass scale and the perturbative
sector provides a consistent description of the neutrino mass
spectrum and mixing angles. The resulting predictions for the
neutrino masses and oscillation parameters, including both the
$m_0 > 0$ and $m_0 < 0$ scenarios, are summarized in
Eq.~(\ref{emppp}). As demonstrated, our model provides a robust
phenomenological fit to the available experimental constraints.

\begin{eqnarray}\label{emppp}
m_1 &=& m_0 \approx \pm(4.87 \text{--} 5.03) \times 10^{-2} \, \text{eV}, \nonumber \\
|m_2| &\approx& \left\{ \begin{matrix} (4.95 \text{--} 5.08) \\ (4.94 \text{--} 5.10) \end{matrix} \right. \text{ or } \left. \begin{matrix} (4.96 \text{--} 5.09) \\ (4.95 \text{--} 5.12) \end{matrix} \right. \times 10^{-2} \, \text{eV}, \nonumber \\
\phi &\approx& \left\{ \begin{matrix} (5.60^{\circ} \text{--} 10.19^{\circ}) \\ (-5.63^{\circ} \text{--} -10.16^{\circ}) \end{matrix} \right. \text{ or } \left. \begin{matrix} (-5.59^{\circ} \text{--} -10.18^{\circ}) \\ (5.61^{\circ} \text{--} 10.12^{\circ}) \end{matrix} \right. , \nonumber \\
m_3 &=& 0, \nonumber \\
\delta m^{2} &\approx& \left\{ \begin{matrix} (5.46 \text{--} 7.86) \\ (6.87 \text{--} 7.10) \end{matrix} \right. \text{ or } \left. \begin{matrix} (6.07 \text{--} 8.85) \\ (7.86 \text{--} 9.13) \end{matrix} \right. \times 10^{-5} \, \text{eV}^{2}, \nonumber \\
|\Delta m^{2}| &\approx& (2.37 \text{--} 2.53) \times 10^{-3} \,
\text{eV}^{2}.
\end{eqnarray}

%=============================================================
\section{Discussion and Phenomenological Implications}\label{sec4}
%=============================================================

\subsection{CP Violation and the Majorana Nature}
A central feature of our framework is the emergence of CP
violation from the symmetry-breaking sector. The phase $\phi$ acts
as the fundamental driver of the Majorana phases. While the global
phase $e^{i\phi}$ is unobservable, the relative phases among the
mass eigenvalues lead to physical consequences in
lepton-number-violating processes.

In the $m_0 < 0$ regime, the model predicts two distinct Majorana
phases: $\rho = (\pi/2 - \phi/2)$ and $\sigma = -\phi/2$.
Interestingly, for the $m_0 > 0$ branch, these phases become
degenerate ($\rho = \sigma = -\phi/2$). This distinction is not
merely mathematical; it defines the effective Majorana mass
$|m_{ee}|$ in neutrinoless double beta decay, providing a
potential experimental probe to distinguish between the two
branches of our model.

\subsection{Cosmological Consistency Check}

To further assess the phenomenological viability of our framework,
we perform a consistency check against the stringent cosmological
bounds on the sum of the neutrino masses. Recent measurements of
the cosmic microwave background by the \textit{Planck} mission
\cite{planck} provide the following upper limit at $95\%$ CL:
\begin{equation}\label{eplank}
\sum m_\nu < 0.12 \, \text{eV} \quad \text{(Planck+WMAP+CMB+BAO)}.
\end{equation}

Within our model, the sum of the neutrino masses is evaluated as
\begin{eqnarray}\label{emsum}
\sum m_\nu &\approx& \begin{cases} (0.0982 \text{--} 0.1011) \,
\text{eV} & \text{for } m_0 > 0, \\ (0.0983 \text{--} 0.1015) \,
\text{eV} & \text{for } m_0 < 0. \end{cases}
\end{eqnarray}
The predicted values lie well within the \textit{Planck} limit,
supporting the phenomenological viability of the TFL symmetric
structure at Point D. Future studies could extend this framework
by investigating its implications for lepton flavor violation
(LFV) processes or by exploring its integration into a more
complete Grand Unified Theory (GUT) scale model.

%=============================================================
%=============================================================
\section{Conclusions}\label{sec5}
%=============================================================
%=============================================================

In this work, we have presented a systematic extension of the
Friedberg-Lee neutrino mass model and established a theoretical
framework in which CP violation emerges naturally from the complex
structure of the mass matrix. Our investigation focused on the
distinctive properties of ``Point D'' at the boundary of the
CP-violating region ($\alpha = \beta = -1/3$). Although Point D
initially gives rise to a Democratic mass matrix characterized by
$S_3$ symmetry and $U_{\text{TBM}}$ mixing, it also leads to a
degenerate mass spectrum that is incompatible with experimental
observations.

At this stage, it is useful to stress the special theoretical role
of Point D in the present framework. This point does not merely
represent a boundary of the CP-violating region, but also defines
the democratic limit from which the physically viable neutrino
mass structure emerges after symmetry breaking. In this sense,
Point D provides the symmetry-based starting point that connects
the underlying $S_3$ structure to the broken TFL scenario studied
in this work.

To remove this degeneracy, we introduced a symmetry-breaking
perturbation that preserves the TFL symmetry at Point D. Unlike
conventional approaches that often rely on \textit{ad hoc}
additions of parameters, our derivation of the perturbation matrix
is guided by the fundamental requirement of identifying the
minimal modification necessary to break both the magic and
$\mu-\tau$ symmetries. This theoretically motivated approach
successfully generates a non-zero Dirac CP phase while preserving
maximal atmospheric mixing, $\theta_{23} = 45^\circ$, and produces
a realistic inverted hierarchy with a massless third-generation
neutrino ($m_3 = 0$).

It is worth emphasizing that the inverted hierarchy obtained in
our framework is a nontrivial consequence of the Twisted
Friedberg--Lee construction. In particular, the democratic limit
at Point D, $\alpha=\beta=-\frac{1}{3}$, leads to a mass spectrum
in which the third neutrino state decouples at leading order,
while the first and second states remain comparatively heavier and
nearly degenerate. This behavior provides a clear physical
distinction between the present model and conventional democratic
scenarios based on $S_3$ symmetry.

The phenomenological viability of the proposed model has been
confirmed through a detailed comparison with the latest
experimental constraints. By determining the allowed parameter
space using the constraints from $\Delta m^2$ and
$\sin\theta_{13}$, we obtained precise predictions for the
neutrino mass scale:
\begin{equation}
m_1 \approx (4.87\text{--}5.03) \times 10^{-2} \, \text{eV}, \quad
|m_2| \approx (4.94\text{--}5.12) \times 10^{-2} \, \text{eV},
\quad \text{and} \quad m_3 = 0.
\end{equation}
Furthermore, our predictions for the leptonic CP-violating
parameters, $\delta \approx
(229.30^{\circ}\text{--}312.42^{\circ})$ and $J \approx
(0.027\text{--}0.036)$, are in excellent agreement with global
neutrino oscillation data and remain fully consistent with the
stringent cosmological bounds reported by the \textit{Planck}
mission.

In summary, by exploiting the distinctive properties of Point D,
we have demonstrated that a minimal and theoretically motivated
perturbation can effectively bridge the gap between highly
symmetric neutrino mass models and experimental observations. The
resulting neutrino mass spectrum and CP-violating phases provide
clear and testable predictions for upcoming neutrino oscillation
experiments. Moreover, the intrinsic CP-violating structure
suggests that the TFL symmetry may serve as an important
low-energy foundation for high-scale scenarios such as
leptogenesis. Future studies incorporating more precise
measurements of the Dirac CP phase and the neutrino mass hierarchy
will be essential to further constrain the parameter space and
strengthen the theoretical foundation of this broken TFL
framework.

%=============================================================
\section{Acknowledgement}
%=============================================================
The authors would like to thank the Research Office of the Qazvin
Branch, Islamic Azad University, for their cooperation.

%=============================================================
\section{Appendix A: The TFL Symmetry and its Algebraic Origin}
\label{app:TFL_symmetry}
%=============================================================
To establish the theoretical underpinning of the phenomenological
results presented in this work, we examine the implementation of
the Twisted Friedberg-Lee (TFL) symmetry \cite{TFL} within the
Dirac neutrino sector. The underlying Lagrangian density for the
Dirac neutrino fields is given by:
\begin{equation}
    -\mathcal{L}_D = \bar{\nu}_{Li} M^D_{ij} \nu_{Rj} + \text{h.c.}
\end{equation}
Within the TFL framework, the neutrino fields $\nu_{L}$ and
$\nu_{R}$ transform independently under the symmetry group,
characterized by the following transformations:
\begin{align}
    \nu_{Li} \longrightarrow S_{ij}^L \nu_{Lj} + \Lambda_{Lj} z, \\
    \nu_{Ri} \longrightarrow S_{ij}^R \nu_{Rj} + \Lambda_{Rj} z,
\end{align}
where $S_{ij}^{L,R}$ denote the permutation matrices and
$\Lambda_{L,R}$ represent the translation parameters.

The requirement of $\mu-\tau$ symmetry further constrains the
parameterization of the Dirac mass matrix $M^D$ to the following
general form:
\begin{equation}
    M^D =
    \begin{pmatrix}
        D & -2C & -2C \\
        -2B & -A & -A \\
        -2B & -A & -A
    \end{pmatrix}.
\end{equation}
The condition of translational invariance necessitates that $M^D$
annihilates the translation vectors $\Lambda_{L,R}$, such that
$M^D \Lambda_{R} = 0$ and $\Lambda_{L} M^D = 0$. By imposing the
case of uniform translation, where $\Lambda_{L,R} \propto (1, 1,
1)$, the algebraic constraints on the matrix elements become fully
determined. Consequently, the Dirac neutrino mass matrix in the
exact symmetry limit simplifies to:
\begin{equation}
    M^D = C
    \begin{pmatrix}
        4 & -2 & -2 \\
        -2 & 1 & 1 \\
        -2 & 1 & 1
    \end{pmatrix}.
\end{equation}
This symmetric structure, characterized by its democratic-like
features, serves as the starting point for the systematic
perturbation analysis developed in the main text.

%=============================================================
\section*{References}
%=============================================================

\end{document}